\documentclass[journal=apchd5,manuscript=article]{achemso}
\usepackage[utf8]{inputenc}
\usepackage{amsmath}
\usepackage{achemso}
\usepackage{siunitx}
\usepackage{amssymb}
\usepackage{verbatim}
\usepackage[normalem]{ulem}
\usepackage{xr}
\usepackage{ulem}
\usepackage{xcolor}
\usepackage[hidelinks]{hyperref}

\usepackage[labelfont=bf]{caption}

\makeatletter

\title[Highly confined, low-loss, and incident-angle-robust surface phonon polariton quasi-bound states in the continuum metasurfaces]{Highly confined incident-angle-robust surface phonon polariton bound states in the continuum metasurfaces}

\author{Lin Nan}
\email{N.Lin@physik.uni-muenchen.de}
\affiliation[Equal]
{These authors contributed equally}
\alsoaffiliation{Chair in Hybrid Nanosystems, Nanoinstitute Munich, Faculty of Physics, Ludwig-Maximilians-Universit\"at M\"unchen, 80539 M\"unchen, Germany}

\author{Andrea Mancini}
\email{A.Mancini@physik.uni-muenchen.de}
\affiliation[Equal]
{These authors contributed equally}
\alsoaffiliation{Chair in Hybrid Nanosystems, Nanoinstitute Munich, Faculty of Physics, Ludwig-Maximilians-Universit\"at M\"unchen, 80539 M\"unchen, Germany}

\author{Thomas Weber}
\affiliation[Equal]
{These authors contributed equally}
\alsoaffiliation{Chair in Hybrid Nanosystems, Nanoinstitute Munich, Faculty of Physics, Ludwig-Maximilians-Universit\"at M\"unchen, 80539 M\"unchen, Germany}

\author{Geok Leng Seah}
\affiliation[NTU]
{School of Materials Science Engineering, Nanyang Technological University, 639798, Singapore, Singapore}

\author{Emiliano Cort\'es}
\affiliation[LMU]
{Chair in Hybrid Nanosystems, Nanoinstitute Munich, Faculty of Physics, Ludwig-Maximilians-Universit\"at M\"unchen, 80539 M\"unchen, Germany}

\author{Andreas Tittl}
\email{Andreas.Tittl@physik.uni-muenchen.de}
\affiliation[LMU]
{Chair in Hybrid Nanosystems, Nanoinstitute Munich, Faculty of Physics, Ludwig-Maximilians-Universit\"at M\"unchen, 80539 M\"unchen, Germany}

\author{Stefan A. Maier}
\affiliation[Australia]
{School of Physics and Astronomy, Monash University, Clayton, Victoria 3800, Australia}
\alsoaffiliation{Department of Physics, Imperial College London, London SW7 2AZ, United Kingdom}

\newcommand*{\addFileDependency}[1]{
\typeout{(#1)}
\@addtofilelist{#1}
\IfFileExists{#1}{}{\typeout{No file #1.}}
}\makeatother

\newcommand*{\myexternaldocument}[1]{%
\externaldocument{#1}%
\addFileDependency{#1.tex}%
\addFileDependency{#1.aux}%
}

\myexternaldocument{Supporting_Info}

\begin{document}

\begin{abstract}


Squeezing light into subwavelength dimensions is vital for on-chip integration of photonic technologies. One approach to overcome the diffraction limit is coupling light to material excitations, leading to polariton states. Here, we showcase how low-loss mid-infrared surface phonon polaritons enable metasurfaces supporting quasi-bound states in the continuum (qBICs) with deeply subwavelength unit cells. Utilizing \SI{100}{\nm} thick free-standing silicon carbide membranes, we achieve highly confined qBIC states with a unit cell volume $\sim 10^4$ times smaller than the diffraction limit. This results in remarkable robustness of the platform against the incident angle that is unique among qBIC systems. We also demonstrate vibrational strong coupling with a thin layer of spin-coated molecules, leveraging the small mode volume. This work introduces phononic qBICs as a novel ultra-confined nanophotonic platform, paving a way for the miniaturization of mid-infrared devices for molecular sensing and thermal radiation engineering.

\end{abstract}

\pagestyle{plain}

\section{Introduction}\label{sec1}

Borrowed from a concept developed in quantum mechanics \cite{von1929some}, photonic bound states in the continuum (BICs) are localized states with formally infinite lifetimes at energies within the radiation continuum \cite{azzam2021photonic,hsu2016BIC}. While true BICs are dark modes, coupling to far-field radiation can be achieved by geometrical detuning of the structure from the BIC state, opening a radiative channel leading to the emergence of so-called quasi-bound states in the continuum (qBICs). Of particular interest are ``symmetry-protected" qBICs within metasurfaces, where coupling to the continuum results from the deliberate symmetry breaking of the unit cell structure. These qBICs boast remarkable attributes, including high and tunable quality factors ($Q$ factor) controlled by the asymmetry parameter \cite{koshelev2018asymmetric,chukhrov2021excitation}, readily adjustable resonance frequency through geometric size scaling \cite{tittl2018imaging}, and  robustness against fabrication errors \cite{kuhne2021fabrication, luorobust}. These compelling attributes have driven diverse applications of symmetry-protected qBICs, including but not limited to mid-infrared vibrational spectroscopy \cite{tittl2018imaging, yesilkoy2019ultrasensitive, aigner2022plasmonic}, refractive index sensing \cite{wang2021all, hsiao2022ultrasensitive}, enhanced light-matter interactions \cite{weber2023intrinsic, sortino2023optically}, photocatalysis \cite{hu2022catalytic}, lasing \cite{kodigala2017lasing, hwang2021ultralow, zhang2022chiral}, and thermal radiation engineering \cite{overvig2021thermal, nolen2023arbitrarily}

The $Q$ factor of qBICs is governed by both the radiation leakage (depending on the degree of asymmetry) and the inherent non-radiative decay due to material absorption. To enhance the $Q$ factor, the simplest strategy is to utilize low-loss dielectrics with negligible intrinsic dissipation \cite{tittl2018imaging, kuhner2023high, sortino2023optically}. Recent endeavors have introduced plasmonic qBICs to amplify surface field enhancement, albeit with reduced $Q$ factors owing to increased losses from electron-electron scattering in noble metals \cite{liang2020bound, aigner2022plasmonic}. Polar dielectrics supporting surface phonon polaritons (SPhPs) can effectively combine the benefits of both material platforms: high field enhancement due to the excitation of surface polaritons in the Reststrahlen (RS) band, and high $Q$ factors due to the slow phonon-phonon scattering mechanism determining their decay \cite{hillenbrand2002phonon,caldwell2013low,caldwell2015low,lee2020image,mancini2020near,tang2022photo}. Furthermore, qBICs in both dielectric and plasmonic metasurfaces typically feature lateral unit cell sizes on the order of the resonance wavelength, hindering their miniaturization \cite{Li2019alldielectric,berger2023,kuhner2022radial}. While deeply sub-diffractional unit cells volumes ($V_\mathrm{diff} = (\lambda/2)^3$ as imposed by the diffraction limit) are in principle feasible in plasmonic qBICs, practical realization is impeded by high losses in the high confinement region where $\operatorname{Re}(\varepsilon)\lesssim0$. Owing to the intrinsic longer SPhPs lifetimes, ``phononic" qBICs can conversely efficiently operate in the sub-wavelength regime at frequencies where $\operatorname{Re}(\varepsilon)\lesssim0$.

In this work, we demonstrate for the first time deeply sub-wavelength SPhPs-driven symmetry-protected qBICs metasurfaces with a minimum unit cell volume that is $10^4$ times smaller than the free-space diffraction limited volume $V_\mathrm{diff}\sim\lambda^3/10$ at the resonant frequency. We fabricate the structures on commercially available large-area (up to few millimeters) free-standing CMOS-compatible silicon carbide (SiC) membranes \cite{mancini2022near, mancini2023multiplication}, where the $Q$ factor is maximized thanks to the symmetric environment provided by the absence of a solid substrate \cite{vertchenko2021near}. We confirm the qBIC character by amplitude and phase resolved maps obtained through transmission scattering-scanning near-field optical microscopy (sSNOM). We further investigate the unique robustness of phononic qBICs against tilt in the illumination angle, which heavily shifts the resonance frequency in dielectric and plasmonic systems.
Additionally, we demonstrate enhanced mid-infrared (IR) light-matter interactions via surface-enhanced IR absorption and vibrational strong coupling between the qBIC resonance and a spin-coated polymer. Polyethylene glycol (PEG) is chosen as the target molecule as it features a fingerprint vibrational mode lying within the SiC RS. Our work introduces phononic qBICs as a novel nanophotonic platform and paves the way for its implementation in mid-IR applications such a as molecular sensing \cite{berte2018sub,li2018boron,bareza2022phonon} and thermal radiation engineering \cite{wang2017phonon,lu2021engineering,lu2020narrowband}.

\section{Results}

\begin{figure}[htp]
\centering
\includegraphics[width=1\textwidth]{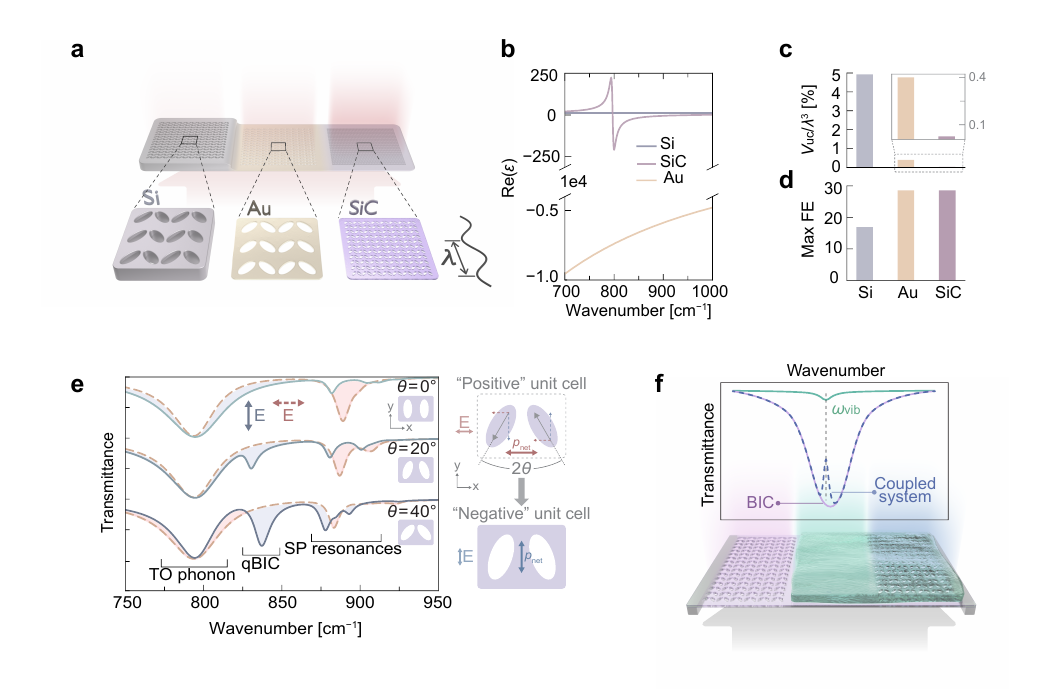}
\caption{\textbf{Highly confined surface phonon polariton driven qBICs.} \textbf{a} Sketch of the unit cell for inverse metasurfaces made of Si (dielectric), Au (metal) and SiC (polar dielectric) showing a qBIC resonance at \SI{11}{\micro \meter}. \textbf{b} Real part of the permittivity for Si, Au, and SiC. Comparison of the unit cell volume \textbf{c} and maximum field enhancement \textbf{d} for the three metasurfaces at the qBIC resonance (the inset in \textbf{c} shows zoomed-in view of only Au and SiC). \textbf{e} Simulated spectra for a SiC ``negative" metasurface with different ellipses tilting angles $\theta$ under excitation with polarization along the $y-$ (blue curves) and $x-$ (red curves) directions. The meaning of ``negative" unit cell is illustrated on the right. \textbf{f} Coating the SiC metasurface with a thin film of molecules having a vibrational mode matching the qBIC frequency induces a splitting of the transmission spectra as a result of enhanced light-matter coupling.}
\label{FigIntro}
\end{figure}

To highlight the advantages brought by our BIC-driven SiC phononic metasurface, we compare it with analogous dielectric and metal structures, namely silicon (Si) and gold (Au) supporting a qBIC resonance at the same wavelength ($\sim$ \SI{11}{\micro \meter}). We consider metasurfaces made of elliptical perforations milled in a film (either Si, Au or SiC) where the radiative channel to enable far-field coupling of the qBIC is activated by the tilting of the ellipses with respect to each other \cite{tittl2018imaging}. The most striking difference between the three materials is in the unit cell volume as shown in Fig. \ref{FigIntro}a, where the SiC structures are remarkably smaller. The real part of the permittivity for the dielectric, plasmonic, and phononic cases is shown in Fig. \ref{FigIntro}b. For the dielectric metasurface (with $\operatorname{Re}(\varepsilon_{\mathrm{Si}}) \sim 10$), the lateral size of the unit cell, although sub-wavelength, is typically comparable to the resonance wavelength, and its minimum thickness is on the order of half a micron. The plasmonic, i.e., gold structure, allows for a thickness reduction by one order of magnitude. However, the lateral size remains the same as the dielectric metasurface because at mid-infrared frequencies noble metals behave almost as perfect electric conductors \cite{novotny2007effective} ($\operatorname{Re}(\varepsilon_{\mathrm{Au}}) \ll 0$). On the other hand, our \SI{100}{nm} thick SiC metasurface supports qBIC resonances even for deeply sub-wavelength unit cells due to the negative, near-zero value of the permittivity ($\operatorname{Re}(\varepsilon_{\mathrm{SiC}}) \lesssim 0$), allowing extreme miniaturization of the device. The quantitative comparison of the unit cell volume obtained from full-three dimensional electromagnetic simulations in  Fig. \ref{FigIntro}c (see Methods for more details) shows that the Au unit cell is an order of magnitude smaller than that of Si, while the SiC unit cell is a further order of magnitude smaller. Moreover, the maximum field enhancement at resonance is approximately twice as large in both the plasmonic and phononic metasurfaces compared to that reached in the dielectric structure (Fig. \ref{FigIntro}d).

Simulated transmission spectra for a SiC metasurface at various ellipses tilting angles $\theta$ are shown in Fig. \ref{FigIntro}e. Each spectrum is computed for two polarizations, aligned with the two orthogonal sides of the unit cell. While a thick SiC layer (several microns) is highly reflective in the RS band \cite{caldwell2013low, mancini2020near}, our \SI{100}{\nm} membranes are below the skin depth and thus highly transparent \cite{mancini2022near}, except for a dip around \SI{800}{\per \cm} due to the TO phonon absorption. In the band from \SI{875}{\per \cm} to \SI{925}{\per \cm}, we observe a series of dips for both polarizations and for all $\theta$, which we associate with modes existing in the single elliptical perforations. At \SI{830}{\per \cm}, we observe a resonance that is associated with a qBIC as it disappears for $\theta=\SI{0}{\degree}$ and exists only for one of the polarizations \cite{tittl2018imaging, kuhne2021fabrication, kuhner2023high}. In conventional ``positive" metasurfaces (i.e. made of antennas placed on a substrate), the net dipole moment, $p_{\mathrm{net}}$, induced by the antennae tilting is along the $x-$direction of the unit cell due to the difference in the gap distances. Consequently, to excite the qBIC, the polarization of the excitation beam has to be along the $x-$direction. According to Babinet's principle, in a ``negative" metasurface (i.e. holes in an extended film), the polarization has to be flipped by \SI{90}{\degree}. Therefore, in our structures, the qBIC is activated when the electric field is along the $y-$direction \cite{berger2023}. 

As a proof-of-concept application of phononic qBICs, we show how our structures can be used to enhance light-matter interactions at mid-IR frequencies. In Fig. \ref{FigIntro}f, we sketch the effect of coating the metasurfaces with a thin layer of molecule with a vibrational mode matching the qBIC resonance. The transmission spectrum of the target molecule, PEG in our case, is represented by a solid green curve denoted as $\omega_{\mathrm{vib}}$, while the spectrum of the bare metasurface is shown in purple. The coupling between the molecule and the metasurface results in a dip in the spectrum, indicated by the dashed curve, which corresponds to the absorption of the vibrational mode. This dip is commonly observed in surface-enhanced infrared absorption (SEIRA) \cite{2014Neubrech,neubrech2017surface,john2023metasurface}. In the case of very efficient light-matter coupling, the regime of strong-coupling can be achieved, resulting in mixed light-matter states where a clear distinction between the two modes cannot be made anymore \cite{torma2014strong, li2018boron, bylinkin2021real,yoo2021ultrastrong, dolado2022remote, weber2023intrinsic}. In the conventional definition, the strong coupling regime is reached when the energy splitting between the two resonances is larger than their individual linewidths.

\begin{figure}[htp]
\centering
\includegraphics[width=1\textwidth]{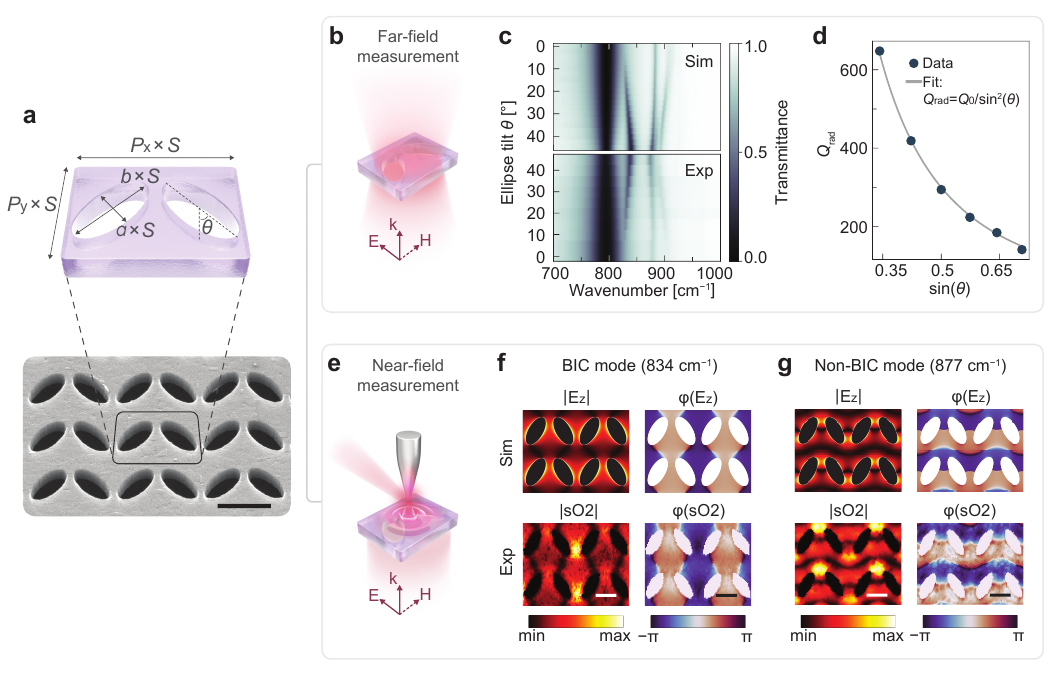}
\caption{\textbf{Far- and near-field characterization of SPhPs-driven qBICs.} \textbf{a} Illustration of the metasurface unit cell with relevant geometrical parameters (top) and tilted SEM image of a portion of a fabricated array (bottom). Scale bar: \SI{500}{nm}. \textbf{b} Sketch of the polarization direction in the transmission FTIR measurements. \textbf{c} Simulated (top) and experimental (bottom) far-field transmission spectra for a metasurface with $S=1$ and varying tilting angle $\theta$. \textbf{d} Radiative $Q$ factor extracted from the TCMT fit of the experimental data in \textbf{c}. \textbf{e} Sketch of the sSNOM setup for the near-field mapping. \textbf{f} Simulated (top) amplitude (left) and phase (right) of the out-of plane component of the surface electric field at the qBIC resonance frequency. The near-field amplitude and phase measured with the sSNOM are shown in the bottom. \textbf{g} Same as in \textbf{f}, but at a frequency of a higher order mode that does not show a qBIC character. Scale bar: \SI{500}{nm}.}
\label{Fig_BIC_demonstr}
\end{figure}

We confirm the realization of qBIC states in SiC metasurfaces through a combination of far-field and near-field experiments. The fabrication process begins with a standard electron-beam lithography step for the creation of a chromium hard mask, followed by a reactive ion etching treatment to mill holes into the membranes (see Methods for more details). A sketch of a single unit cell with relevant geometrical parameters is shown in Fig. \ref{Fig_BIC_demonstr}a together with a SEM image of a portion of one of the fabricated arrays. The long and short axes of the ellipses are set to be $a = \SI{1190}{nm}$ and $b=\SI{560}{nm}$, while the edges of the unit cell are $P_\mathrm{x} = \SI{2100}{nm}$ and $P_\mathrm{y} = \SI{1610}{nm}$. These dimensions are defined as the scaling factor $S=1$. 
While adjusting the thickness of the material offers a means to tune the position of the qBIC resonance frequency (see Supporting Information \ref{SiC_200nm}), altering the scaling factor is a more convenient and efficient method for tuning the resonance of qBIC metasurfaces.\cite{tittl2018imaging, kuhner2023high}. The theoretical spectra shown in Fig. \ref{FigIntro}e is validated through transmittance measurement on our structures with Fourier-transform infrared (FTIR) microscopy (see Methods section for more details). To demonstrate the existence of the qBIC resonance, we maintain all geometrical parameters of the unit cell constant but solely vary the tilting angle $\theta$ between the elliptical holes. Following Babinet's principle, we polarize the electric field of the incident beam along the short axis of the unit cell \cite{berger2023} as sketched in Fig. \ref{Fig_BIC_demonstr}b. When changing the ellipse tilting angle $\theta$, we observe a series of dips in both simulated (top) and experimental (bottom) transmission spectra in Fig. \ref{Fig_BIC_demonstr}c. The lowest-energy dip is associated with the TO-phonon absorption from the bare SiC membrane. At higher energy, the qBIC resonance can be identified as it becomes optically active only for $\theta>\SI{0}{\degree}$. As expected, we observe an increase of the modulation depth and linewidth of the resonance with increasing $\theta$ due to the strengthening of the radiative dissipation channel introduced by the structure asymmetry. Higher-energy dips are associated with higher order resonances of the elliptical meta-atoms as they are present even at $\theta = \SI{0}{\degree}$. The close match between experimental and simulated spectra confirms the accuracy of our theoretical model and interpretation. Interestingly, we observe a narrowing of the TO phonon band-width with increasing $\theta$ and qBIC modulation depth (see Supporting Information \ref{Unit_cell_modulation}), indicating the presence of coupling between the two.

To further confirm the qBIC characteristic of the resonance, we fit the experimental spectra with a temporal coupled mode theory (TCMT) model (for more details see Supporting Information \ref{TCMT}). From this fit, we can extract the radiative part of the resonance $Q$ factor $Q_{\mathrm{rad}}$ as shown in Fig. \ref{Fig_BIC_demonstr}d. As a general feature of symmetry-protected BIC-driven metasurfaces, $Q_{\mathrm{rad}}$ is proportional to the inverse square of the asymmetry parameter, which for a metasurface made of elliptical antennas can be defined as $\sin^2 (\theta)$ \cite{koshelev2018asymmetric}. The $Q_{\mathrm{rad}}$ values extracted from the TCMT model fits well with the $1/\sin^2 (\theta)$ function, confirming the qBIC character of the resonance.

To study the field-enhancement distribution and ultimately confirm the qBIC nature of the observed resonance, we carry out nanoscale mapping with a scattering scanning near-field optical microscope (sSNOM) as sketched in Fig. \ref{Fig_BIC_demonstr}e. We perform the measurements in transmission mode, which is ideal for the investigation of resonators due to reduced tip-sample coupling and normal-incidence illumination \cite{schnell2011nanofocusing, schnell2010phase}. Briefly, a mid-infrared beam from a tunable mid-IR laser source is focused from below by a parabolic mirror on the tip of an atomic force microscope (AFM). The tip scatters the incident light, which is collected by an off-axis upper parabolic mirror. The signal is demodulated at higher harmonics $n\geq2$ of the tip oscillation frequency for far-field background suppression and is analyzed with a pseudo-heterodyne interferometer for extraction of the amplitude and phase response (more details in the Methods). 

For the near-field mapping, we use a metasurface with $S=1$ and $\theta = \SI{30}{\degree}$, of which the far-field response is included in Fig. \ref{Fig_BIC_demonstr}c. Results at frequencies where the structure supports a qBIC ($\omega = \SI{834}{\per \cm}$) and a single elliptical hole resonance ($\omega = \SI{877}{\per \cm}$) are reported in Fig. \ref{Fig_BIC_demonstr}f and g respectively. For each frequency, we show in the top row the simulated amplitude (left) and phase (right) of the out-of plane component of the field $E_z$ at the SiC surface. The corresponding near-field data (demodulation order $n=2$) are shown in the bottom row, demonstrating good agreement between simulated and experimental maps. A small discrepancy can be seen in the experimental amplitude maps, where high field hot-spots appear at the antenna gap that have no counterpart in the simulation. We attribute this effect to the combination of non-perfect normal illumination and residual tip-sample coupling which can impact the measured amplitude, but have negligible effect on the phase profile. We use the topography to mask the optical data in the region of the perforations, where the mechanical tip-sample interaction is unreliable due to the absence of a substrate. Interestingly, we can clearly distinguish the qBIC resonance in Fig. \ref{Fig_BIC_demonstr}f as the phase jumps are in the $x-$direction, which is perpendicular to the polarization of the exciting beam. For the non-qBIC mode shown in Fig. \ref{Fig_BIC_demonstr}g, the phase jumps are, as expected for a dipolar-type mode, along the polarization direction (in the $y-$direction). The combination of far-field and near-field measurements shown in Fig. \ref{Fig_BIC_demonstr} allows us to characterize and unequivocally confirm the qBIC nature of the deeply sub-wavelength resonances we engineer in SiC phononic metasurfaces. 

\begin{figure}[htp]
\centering
\includegraphics[width=0.5\textwidth]{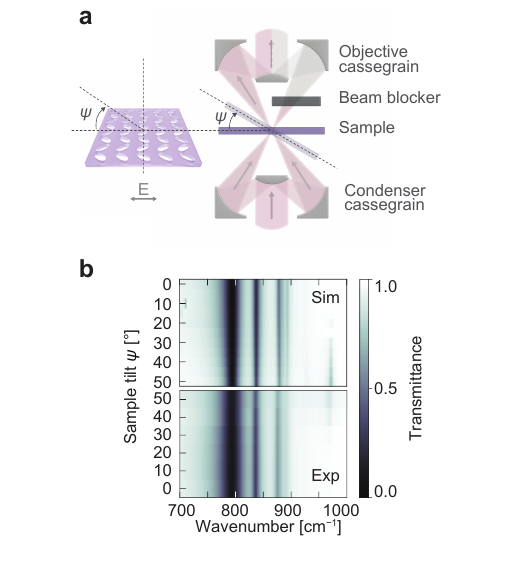}
\caption{\textbf{Angle-robustness of deeply subwavelength qBICs.} \textbf{a} Experimental setup for transmission measurement with various sample tilting angles $\psi$. \textbf{b} Simulated and experimental spectra for a metasurface with $S=1$ and $\theta=\SI{40}{\degree}$ under effectively tilted illumination at multiple angles from normal incidence to $\psi=\SI{50}{\degree}$.}
\label{Fig_BIC_angl_ind}
\end{figure}

In conventional qBIC systems (both plasmonic and dielectric), the resonance position is heavily dependent on the excitation beam incident angle \cite{leitis2019angle, berger2023}. For this reason, refractive objectives with low numerical apertures are generally required to obtain good modulation depths in qBIC systems. A standout feature of our phononic qBIC is its dispersionless nature, which can already be indirectly inferred from the pronounced resonance dips we observe in our far-field spectra (see e.g. Fig. \ref{Fig_BIC_demonstr}c) obtained with off-axis illumination from a Cassegrain objective, the use of which generally degrades the measured resonance modulation \cite{berger2023}. To confirm this observation, we mount our samples on a tilting stage allowing us to measure transmission spectra for different incident angles with respect to the sample normal direction as shown in Fig. \ref{Fig_BIC_angl_ind}a. To ensure that we probe the metasurfaces response with a spread of angles averaged over a single value, we block half of the upper objective. Strikingly, simulations and measurements for a metasurface with $S=1$ and $\theta = \SI{40}{\degree}$ demonstrate the clear independence of all the modes to the illumination angle $\psi$ (Fig. \ref{Fig_BIC_angl_ind} b). Especially, the consistent qBIC resonance positions and their unvarying modulation depths underscore the superior angle-robustness of our system. The only appreciable variation of the spectra can be seen around \SI{970}{\per \cm} due to the excitation of the so-called Brewster mode at the epsilon-near-zero point \cite{taliercio2014brewster}. This interpretation is confirmed by the appearance of this transmission dip only for TM-polarized excitation (for more details see Supporting Information \ref{TM_TE}). We attribute the angle-robustness of our design to the uniquely deeply sub-wavelength character of the phononic qBIC, which implies neighboring unit cells are illuminated with an almost identical phase even for highly tilted illumination. We numerically confirm that a positive phononic qBIC metasurface is also angle independent, while an analogous metallic metasurface is strongly dispersive both in the positive and negative configurations (see Supporting Information \ref{Au_SiC_sim_angle}). This observation excludes the attribution of the angle-independence to the particular metasurface geometry and points to the generality of the phenomena for sub-wavelength unit cells. 

\begin{figure}[htp]
\centering
\includegraphics[width=1\textwidth]{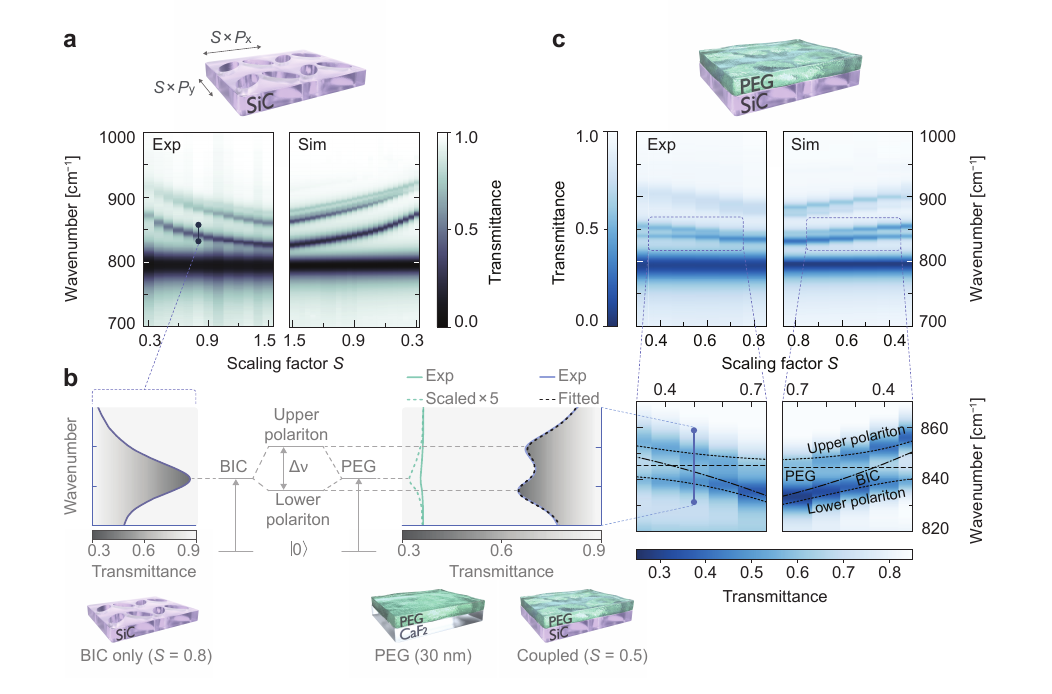}
\caption{\textbf{Vibrational strong coupling of SiC qBICs with PEG.} \textbf{a} Experimental and simulated transmission spectra of bare SiC metasurfaces with scaling factors ranging from 0.4 to 1.5. \textbf{b} Schematic of vibrational strong coupling. The BIC resonance is spectrally overlapped with the absorption band of PEG, resulting in a splitting of the BIC mode into upper and lower polaritonic states. \textbf{c} Experimentally and simulation resolved anti-crossing of with a thickness of \SI{30}{\nano\meter} PEG covered SiC metasurfaces. The dotted lines denote the polariton dispersion calculated with extracted parameters from TCMT fitting and a Rabi splitting of \SI{11.5}{\per\cm}.}
\label{Fig_BIC_PEG}
\end{figure}

The unique combination of deeply sub-diffractional unit cell size and high field enhancement of our phononic qBICs can be exploited to boost light-matter interactions in the mid-IR. As the near-field coupling strength between photonic modes and material excitations is inversely proportional to the square root of the mode volume \cite{chikkaraddy2016single}, qBIC metasurfaces with extremely small effective mode volumes ($V_\mathrm{eff}$) are an ideal platform for highly efficient light-matter interactions. We select polyethylene glycol (PEG) as a prototype molecule that features a characteristic C-H bending mode at $\omega =\SI{842}{\per \cm}$ matching SiC's RS band. PEG is a technologically relevant water-soluble polyether compound widely employed in pharmaceutical, cosmetics and manufacturing industries \cite{d2016polyethylene}. In order to match the qBIC and PEG resonance frequencies, we study the tunability of our metasurfaces via scaling factor tuning, where all the unit cell geometrical parameters are scaled by a single multiplicative factor $S$. As the phononic qBIC resonance can only exist inside the material RS band, we observe in Fig. \ref{Fig_BIC_PEG}a a highly nonlinear frequency shift in the experimental (left) and simulated (right) transmission spectra of a metasurface with $\theta = \SI{40}{\degree}$. The qBIC resonance positions can only asymptotically approach the TO-phonon even for large $S$ values and analogously are bounded from above by the LO-phonon energy. This second effect is not addressed in our experimental study as saturation of the qBIC resonance position for decreasing $S$ is appreciable only for very small unit cell sizes below the resolution limit of standard nanofabrication methods (see Supporting Information \ref{small_UC_limit}). The smallest metasurface fabricated for this study corresponds to $S=0.25$. Importantly, the qBIC tunability extends both above and below the PEG vibrational mode, allowing a comprehensive investigation of the light-matter coupling strength. 

To enable effective near-field interaction, we cover our SiC metasurfaces with a layer of PEG with an estimated thickness of \SI{30}{\nm} by spin coating (see Supporting Information \ref{PEG_thickness}). The effect of the PEG layer on the qBIC resonance is shown in Fig. \ref{Fig_BIC_PEG}b. When the frequencies of both modes overlap, a splitting of the qBIC dip can be observed in the transmission spectra, corresponding to the creation of two polariton branches of different energies. For numerical modeling and fitting of the coupled states, the dielectric function of PEG was calculated from Kramers-Kronig consistent fits to transmission measurements of films of different thicknesses on a CaF$_2$ substrate. The extracted linewidth of the PEG absorption band was $\gamma_\mathrm{PEG} = \SI{3.59}{\per\cm}$, which we fix for the TCMT fits used to extract the coupling strength. In Fig. \ref{Fig_BIC_PEG}c, we report experimental (left) and simulated (right) transmission spectra for the coupled system when varying the metasurface scaling factor. Two changes in the optical response of our SiC metasurface can be observed: (i) a spectral red-shift of the qBIC resonances due to the change in the local refractive index of the environment, resulting in a reduction of the scaling factor needed to achieve spectral overlap between qBIC and PEG, and (ii) a band-splitting at the absorption peak of PEG around \SI{842}{\per \cm} into upper and lower polaritonic branches as illustrated in Fig. \ref{Fig_BIC_PEG}b. Significantly, the absorption fingerprint of the molecule on the metasurfaces is strongly enhanced in comparison to the bare CaF$_2$ substrates, which can be attributed to the pronounced near-field enhancement of the qBIC mode. Detuning the qBIC mode from the PEG absorption by varying the scaling factor $S$ reveals the typical anti-crossing pattern characteristic of strongly coupled resonances. In the numerical simulations we fill the elliptical holes of the membrane completely and add a \SI{30}{\nm} thick layer on the top side of the membrane, resulting in excellent agreement with our experimental results. To quantify the strength of the vibrational coupling, we fit transmittance spectra with our TCMT model (for more information on the fitting procedure see Supporting Information \ref{TCMT_vibrational_coupling}), which allows for the extraction of the dispersion and linewidth of the qBIC from coupled spectra, which would be unattainable from mere polariton dispersion fits.

As can be seen in Fig. \ref{Fig_SI_TCMT_cmp_res_pos_fits}, the extracted qBIC dispersion from simulations follows a similar quadratic and continuous dependence on the scaling factor that we observe from experimental measurements of the bare metasurfaces. Because the thickness of the PEG varies slightly depending on the scaling factor in experiment, we instead approximate the qBIC dispersion with a linear relationship. Furthermore, we simplify the system by assuming a constant qBIC linewidth over the sampled scaling factor range, which is consistent to commonly utilized fits to extract the Rabi splitting of anti-crossing spectra. The damping rate of the qBIC is extracted to be $\gamma_\mathrm{BIC} = \SI{7.57}{\per\cm}$ and the near-field coupling strength at the point of spectral overlap between qBIC and PEG to be $g = \SI{6.08}{\per\cm}$ via interpolation at a scaling factor of $S=0.435$. In order for the system to be in the strong-coupling regime, the exchange of energy between BIC and PEG needs to be faster than the constituting resonances can dissipate energy into their individual loss channels \cite{chikkaraddy2016single, weber2023intrinsic}. In other words, the frequency separation of the polariton peaks (the vacuum Rabi splitting $2g$) must be larger than the sum of the linewidths of BIC and PEG. This is commonly referred as the strong coupling criterion $c_1 = 2g/(\gamma_\mathrm{BIC} + \gamma_\mathrm{PEG}) > 1$, which we evaluate for our system to be $c_1 = 1.09$. Because of the mismatch of the BIC and PEG linewidths, the peak splitting is actually given by the Rabi splitting $\Omega_\mathrm{R} = 2\sqrt{g^2 - (\gamma_\mathrm{BIC} - \gamma_\mathrm{PEG})^2 / 4} = \SI{11.5}{\per\cm}$, which is slightly lower than $2g$ and a stricter criterion is given by $c_2 = \Omega_\mathrm{R}/(\gamma_\mathrm{BIC} + \gamma_\mathrm{PEG}) > 1$ with $c_2 = 1.03$. Both criteria are fulfilled and show that we reach the strong coupling regime. Finally, we want to remark that thanks to the low mode volume of phononic qBICs we are able to reach the strong-coupling regime even with the relatively weak vibrational peak of PEG when compared with molecules routinely used for mid-IR vibrational strong-coupling such as SiO$_2$ and PMMA \cite{bylinkin2021real, yoo2021ultrastrong, dolado2022remote, PhysRevLett.131.126902} (see Supporting Information \ref{molecules_stronger_coupling}). These observations highlight the potential of phononic qBICs for the investigation and exploitation of mid-IR strong coupling at the nanoscale.

\section{Conclusions}
In this work, we reported the pioneering realization of a phonon polariton-mediated qBIC metasurface. The qBIC nature was unequivocally assessed through comprehensive far- and near-field characterizations, supported by rigorous numerical simulations. By utilizing a highly dispersive low-loss polaritonic material, we achieved extreme miniaturization by shrinking the unit cell size down to $\sim1.5\cdot 10^{-4}$ of the volume $\lambda_0^3/10$ occupied by the free-space wavelength. This remarkable miniaturization led to the observation of a striking incident-angle invariant behavior on our metasurfaces, a highly unconventional characteristic for a symmetry-protected BIC metasurface \cite{leitis2019angle, berger2023}. This result paves a way for the utilization of phononic qBIC metasurfaces, and sub-wavelength polaritonic systems in general, in applications requiring consistent performances over a wide range of incident angles such as in stealth technologies \cite{zhong2017radar} and omnidirectional radiative cooling \cite{zou2017metal, lin2023highly}. Moreover, the strong confinement of the mode volume facilitated the emergence of vibrational strong coupling between a weak resonator, exemplified by PEG, and our qBIC system. While phonon polariton vibrational strong coupling has been shown with hBN antennas obtained from exfoliated flakes \cite{li2018boron}, we report here its first realization in a large-area CMOS-compatible platform. The use of SiC is therefore suited for direct applications in SEIRA \cite{neubrech2017surface, john2023metasurface}, nonlinear mid-IR optics \cite{razdolski2018second, kitade2020nonlinear} and for the control of chemical-reaction pathways through strong-coupling engineering \cite{thomas2019tilting, garcia2021manipulating} at the nanoscale.

\section*{Methods}\label{sec11}

\subsection*{Sample preparation}
Metasurfaces were fabricated on free-standing \SI{100}{nm} SiC membranes supported by Si frame that we purchased from Silson Ltd. Fabrication follows the conventional top-down nanofabrication method. First, the electron-beam resist,  SX AR-N 8200.06, was exposed with \SI{10}{micron} aperture under \SI{30}{KV} acceleration voltage, followed by the deposition of chromium as a hard-mask for reactive ion etching (RIE) via electron-beam deposition. SF$_{\mathrm{6}}$ based RIE was performed to perforate the SiC membrane followed by wet-etching process for the removal of the hard-mask. 

For the vibrational coupling measurements, 2000K PEG was purchased from Sigma aldrich and dissolved into methoxybenzene to achieve \SI{12.5}{mg/mL}. The PEG solution was spin-coated onto the suspended SiC metasurfaces. The PEG thickness on SiC metasurfaces was estimated by interpolating the thickness-absorption correlated curve acquired from ellipsometry and Fourier transform infrared (FTIR) spectroscopy on bulk substrates (for more detials, see Supporting Information \ref{PEG_thickness}.)

\subsection*{Fourier Transform Infrared spectroscopy}
Transmission spectra of the metasurfaces were acquired using a VERTEX 80v FTIR spectrometer (Bruker) in conjunction with a HYPERION 3000 microscope (Bruker). Measurements were performed utilizing a reflective microscope objective (Newport), also known as Cassegrain objective, featuring a numerical aperture (NA) of 0.4 and 15× magnification covering a range of polar angles from \SI{12}{\degree} to \SI{23.6}{\degree}. A liquid nitrogen-cooled mercury cadmium telluride detector recorded the spectra in transmission mode. The sample tilting was achieved by mounting the sample on a custom-made sample holder that can be rotated along central axis that is perpendicular to the effective beam path. Custom-made aperture was added onto the objective over the half of the transmitted beam for the tilted measurements.

\subsection*{Transmission sSNOM}

Near-field spectra were obtained with a commercial sSNOM set-up (Neaspec) equipped with a pseudo-heterodyne interferometer to obtain amplitude and phase resolved images. The light source used in the experiments was an optical parametric oscillator (OPO) laser (Stuttgart Instruments) powered by a pump laser at $\lambda=\SI{1035}{\nano \meter}$ with $\SI{40}{\mega \hertz}$ repetition rate and $\approx\SI{500}{\femto \second}$ pulses. The linearly polarized MIR output was obtained by difference frequency generation in a nonlinear crystal between the signal and idler outputs of the OPO. The beam frequency bandwidth was reduced using a monochromator. The beam was loosely focused at normal incidence by a parabolic mirror positioned below the sample. The near-field signal was scattered by a metal-coated (Pt/Ir) atomic force microscope tip (Arrow-NCPt, Nanoworld) oscillating at a frequency $\Omega\approx\SI{280}{\kilo \hertz}$, and collected by a second off-axis parabolic mirror positioned above the sample. The tapping amplitude was set to $\approx\SI{80}{\nano\meter}$ and the signal demodulated at the second harmonic $2\Omega$ for background suppression. Before focusing, half of the light was redirected towards a pseudo-heterodyne interferometer used to retrieve both amplitude and phase of the signal. The light scattered by the tip was recombined with the interferometer reference arm by a second beam-splitter and directed towards a nitrogen-cooled mercury cadmium telluride infrared detector.

\subsection*{Simulations}

Electromagnetic simulations were performed with a commercial solver (CST Studio) in frequency domain. The dielectric function of SiC was modeled as follows:

\begin{equation}
    \varepsilon(\omega) = \varepsilon_{\infty} \left( 1+ \frac{\omega_{LO}^2 - \omega_{TO}^2}{\omega_{TO}^2 - \omega^2 - i\gamma \omega}\right)
\end{equation}

with $\omega_{TO} = \SI{797}{\per \cm}$, $\omega_{LO} = \SI{973}{\per \cm}$, $\varepsilon_{\infty}=6.6$ and $\gamma = \SI{6.6}{\per \cm}$. 
The PEG dielectric function was modeled with a single Lorentzian as:

\begin{equation}
    \varepsilon_\mathrm{PEG}(\omega) = \varepsilon_{\mathrm{PEG},\infty} + \frac{A_\mathrm{PEG}}{\omega_{\mathrm{PEG},0}^2 - \omega^2 - 2i\gamma_\mathrm{PEG}\omega}
\end{equation}

where $\varepsilon_{\mathrm{PEG},\infty}$ is set to 2.25, $\omega_{\mathrm{PEG},0} = \SI{842.4}{\per\cm}$, $\gamma_\mathrm{PEG} = \SI{3.59}{\per\cm}$ and the oscillator strength $A_\mathrm{PEG} = \SI{8026}{\per\cm\squared}$. The factor of two in the denominator is due to the convention chosen to be comparable to our TCMT model, where $\gamma$ refers to the half width at half maximum of a Lorentzian.

The metasurface response was simulated by using periodic Floquet boundary conditions and plane-wave excitation. The electric field distributions were extracted at the top layer of the SiC film.

\subsection*{Acknowledgments}

This project was funded by the Deutsche Forschungsgemeinschaft (DFG, German Research Foundation) under grant numbers EXC 2089/1–390776260 (Germany’s Excellence Strategy) and TI 1063/1 (Emmy Noether Program), the Bavarian program Solar Energies Go Hybrid (SolTech) and the Center for NanoScience (CeNS). Funded by the European Union (METANEXT, 101078018 and NEHO, 101046329). L.N. and E.C acknowledge funding from the European Commission through the ERC Starting Grant CATALIGHT (802989) and DAAD (57573042). Views and opinions expressed are however those of the author(s) only and do not necessarily reflect those of the European Union or the European Research Council Executive Agency. Neither the European Union nor the granting authority can be held responsible for them. S.A.M. additionally acknowledges the Lee-Lucas Chair in Physics and the EPSRC (EP/W017075/1). 

\bibliography{sn-bibliography.bib}

\end{document}